\documentclass[11pt]{amsart}
\usepackage{geometry}                % See geometry.pdf to learn the layout options. There are lots.
\usepackage{graphicx}
\usepackage{amssymb}
\usepackage{epstopdf}
\usepackage{array}
\usepackage{subfigure}
\usepackage{url}

\title[Physarum wires]{Physarum wires: \\ Self-growing self-repairing smart wires \\ made from slime mould}
\author[Adamatzky]{{\bf Andrew Adamatzky}\\ \vspace{0.1cm}  \\ Unconventional Computing Centre, \\ University of the West of England,\\ Bristol, BS16 1QY, United Kingdom}
\address[Adamatzky]{University of the West of England, Bristol, BS 16 1QY, United Kingdom}

\begin{document}

\maketitle

\begin{abstract}

We report experimental laboratory studies on developing conductive pathways, or wires, using protoplasmic tubes of 
plasmodium of acellular slime mould \emph{Physarum polycephalum}.  Given two pins to be connected by  a wire, we place a 
piece of slime mould on one pin and an attractant on another pin. Physarum propagates towards the attract and thus connects the pins with 
a protoplasmic tube. A protoplasmic tube is conductive, can survive substantial over-voltage and can be used to transfer electrical current 
to lightning and actuating devices. In experiments we show how to route Physarum wires with  chemoattractants and electrical fields.  We demonstrate
that Physarum wire can be grown on almost bare breadboards and on top of electronic circuits. The Physarum wires can be insulated with a 
silicon oil without loss of functionality. We show that a Physarum wire self-heals: end of a cut wire merge together and restore the conductive pathway in several hours after being cut. Results presented will be used in future designs of self-growing wetware circuits and devices, and integration of 
slime mould electronics into unconventional bio-hybrid systems.  

\end{abstract}

\section{Introduction}

The plasmodium of \emph{Physarum polycephalum} (Order \emph{Physarales}, class \emph{Myxomecetes}, subclass \emph{Myxogastromycetidae}) is a single cell, visible with the naked eye, with many diploid nuclei. The plasmodium feeds on bacteria and microscopic food particles by endocytosis. When placed in an environment with distributed sources of nutrients the plasmodium forms a network of protoplasmic tubes connecting the food sources. In 2000 Nakagaki et al \cite{nakagaki_2000} showed that the topology of the plasmodium's protoplasmic network optimizes the plasmodium's harvesting of nutrient resource from the scattered sources of nutrients and makes more efficient the transport of intra- cellular components. 
In \cite{adamatzky_physarummachines} we have shown how to construct specialised and general purpose massively-parallel amorphous computers from the plasmodium (slime mould) of \emph{P. polycephalum} that are capable of solving problems of computational geometry, graph-theory and logic.

Plasmodium's foraging behaviour can be  interpreted as a computation~\cite{nakagaki_2000,nakagaki_2001a,nakagaki_iima_2007}:  data are represented by spatial of attractants and repellents, and  results are represented by structure of protoplasmic  
network~\cite{adamatzky_physarummachines}.  
Plasmodium can solve computational problems with natural parallelism, e.g. related to shortest 
path~\cite{nakagaki_2001a} and hierarchies of planar proximity graphs~\cite{adamatzky_ppl_2008}, computation of plane tessellations~\cite{shirakawa}, execution of logical computing schemes~\cite{tsuda2004,adamatzky_gates}, and natural implementation of spatial logic and process algebra~\cite{schumann_adamatzky_2009}.  

In the framework of our ``Physarum Chip'' EU project~\cite{adamatzky_phychip} we aim to experimentally implement a working prototype of 
a Physarum based general purpose computer. This computer will combine self-growing computing circuits made of a living slime mould with conventional electronic components. Data and control inputs to the Physarum Chip will be implemented via chemical, mechanical and optical means.

 Aiming to develop a component base of future Physarum computers we designed Physarum tactile sensor~\cite{adamatzky_2013_tactile} and
 undertook foundational studies towards fabrication of  slime mould chemical sensors (Physarum nose)~\cite{delacycostello_2013, whiting_2013}.
 We have uncovered memristive properties of the slime mould~\cite{gale_2013}: Physarum memristors could be a basic (logically universal) components of future slime mould computers.
 
 To transfer data, elements of a program code and control signals in Physarum computers we must develop a reliable self-growing and self-healing conductive pathways made of the slime mould --- Physarum wires. 
 
 In present paper  we outline our scoping experimental results on evaluating properties of protoplasmic tubes as electrical wires. Our finding 
 discussed in the paper complement previous studies on fabrication of organic wires and using living substrates to grow metallised conductive pathways: 
self-assembling molecular wires~\cite{wang_2006,paul_1998}, DNA wires~\cite{beratan_1997}, 
electron transfer pathways in biological systems~\cite{katz_2005}, live bacteria templates for conductive 
pathways~\cite{berry_2005},  bio-wires with cardiac tissues~\cite{cingolani_2012}, golden wires with templates 
of fungi~\cite{sabah_2012}. 

The paper is structured as follows. We outline experimental techniques in Sect.~\ref{methods}. 
Section~\ref{resistivity_overload_transfer} presents experimental results on resistivity of Physarum wires, behaviour of the wires
under electrical load and potential division. In Sect.~\ref{routing} we show how to route Physarum wires with chemo-attractants, 
chemo-repellents and electrical fields. An ability of Physarum wires to restore their integrity after damage is experimentally 
proved in Sect.~\ref{self_repair}.  Section~\ref{electronic_boards} exemplifies growth of Physarum wires on electronic boards. 
Insulation of Physarum wires is studied in Sect.~\ref{insulating}. Section~\ref{discussion} discusses pros and cons of conductive pathways 
made from Physarum's protoplasmic tubes.

\section{Methods}
\label{methods}

\begin{figure}[!tbp]
\centering
\includegraphics[width=0.8\textwidth]{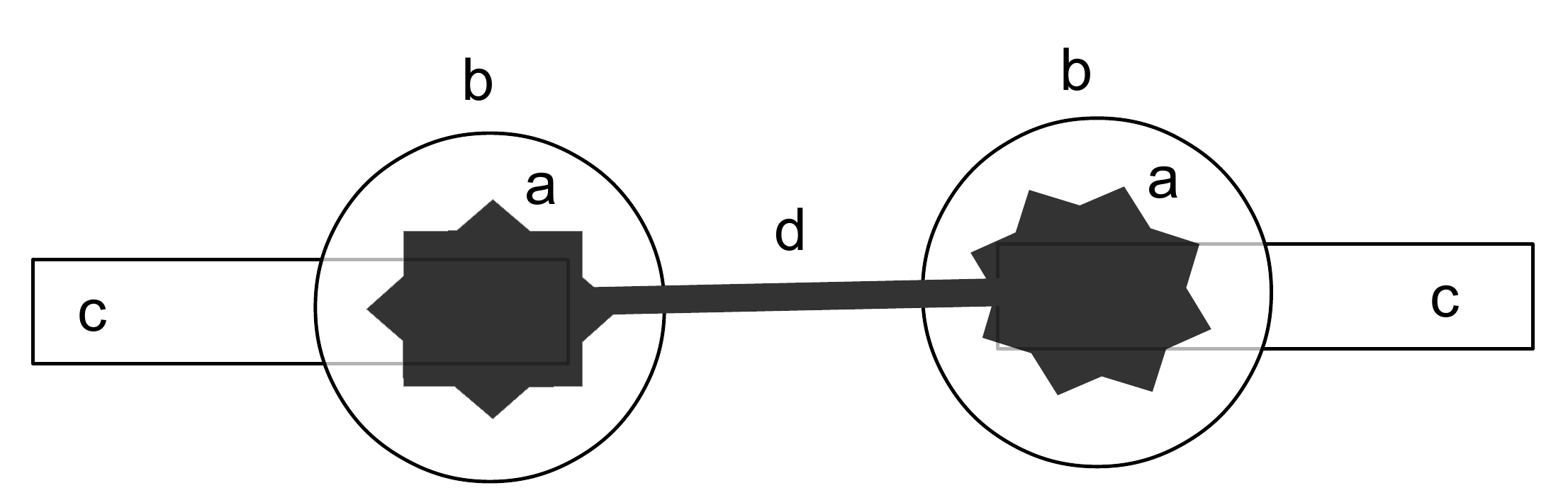}\\
\includegraphics[width=0.8\textwidth]{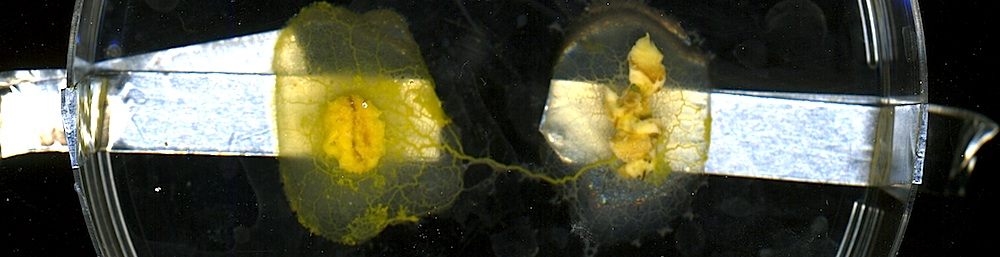}\\
\caption{Top: A scheme of experimental setup. (a)~Physarum, (b)~agar blobs, (c)~electrodes, (d)~protoplasmic tube. All parts of Physarum 
shown in dark grey form a single cell. Bottom: A snapshot of agar blobs occupied by Physarum and connected by a protoplasmic tube.}
\label{scheme}
\end{figure}

Plasmodium of \emph{Physarum polycephalum} was cultivated in plastic lunch boxes (with few holes punched in their lids for ventilation) on wet kitchen towels and fed with oat flakes. Culture was periodically replanted to a fresh substrate. Electrical activity of plasmodium was recorded with  
ADC-24 High Resolution Data Logger  (Pico Technology, UK).  A scheme of experimental setup is shown in Fig.~\ref{scheme}. Two blobs of agar 
2 ml each (Fig.~\ref{scheme}b) were placed on electrodes (Fig.~\ref{scheme}c) stuck  to a bottom of a plastic Petri dish (9~cm). Distance between proximal sites of electrodes is always 10~mm. Physarum was inoculated on one agar blob. We waited till Physarum colonised the first blob, where it was inoculated, and propagated towards and colonised the second blob. When second blob is colonised, two blobs of agar, both colonised by Physarum (Fig.~\ref{scheme}a), became connected by a single protoplasmic tube  (Fig.~\ref{scheme}d). Resistivity and voltage over 2.5~V were measured on TTi 1604 Digital Multimeter; Physarum extracellular potential was measured using PicoTech ADC-20/24 logger. Voltage and current supplied to hybrid circuits incorporating Physarum wires was supplied using Iso-Tech IPS 4303 laboratory DC power supply unit. When growing Physarum on breadboard we made a layer of agar on one side of the board, to keep humidity higher, and  inoculated Physarum on other side of the board. After Physarum reached its destination we removed the agar layer, thus no  conductive substance apart of slime mould present on the breadboard. Images of Physarum  were made using Fuji FinePix S6500 camera, EpsonPerfection 4490 scanner and DinoLite Computer Microscope.

\section{Resistivity, overload and transfer function}
\label{resistivity_overload_transfer}

In 25 experiments we measured resistance and calculated resistivity of protoplasmic tubes on agar blobs. In calculations we assumed length 
of a tube is 1~cm, and diameter is 0.03~cm. We found that minimum resistance recorded is 80~$\Omega$$\cdot$cm, 
maximum resistance is 2560~$\Omega$$\cdot$cm,  median 625~$\Omega$$\cdot$cm, and average 825~$\Omega$$\cdot$cm.  Resistivity of Physarum substantially varies from one experiment to another: standard deviation calculated is 776, which is just slightly below average of Physarum wire resistivity resistivity. Average resistivity of Physarum protoplasmic tubes is of the same rank as resistivity of a cardiac muscle of a dog, and skeletal muscles of a dog and a human~\cite{geddes_baker_1967}.
%Note, that resistivity of an agar blob is usually 400~$\Omega$$\cdot$cm. 

\begin{figure}[!tbp]
\centering
\subfigure[]{\includegraphics[width=0.49\textwidth]{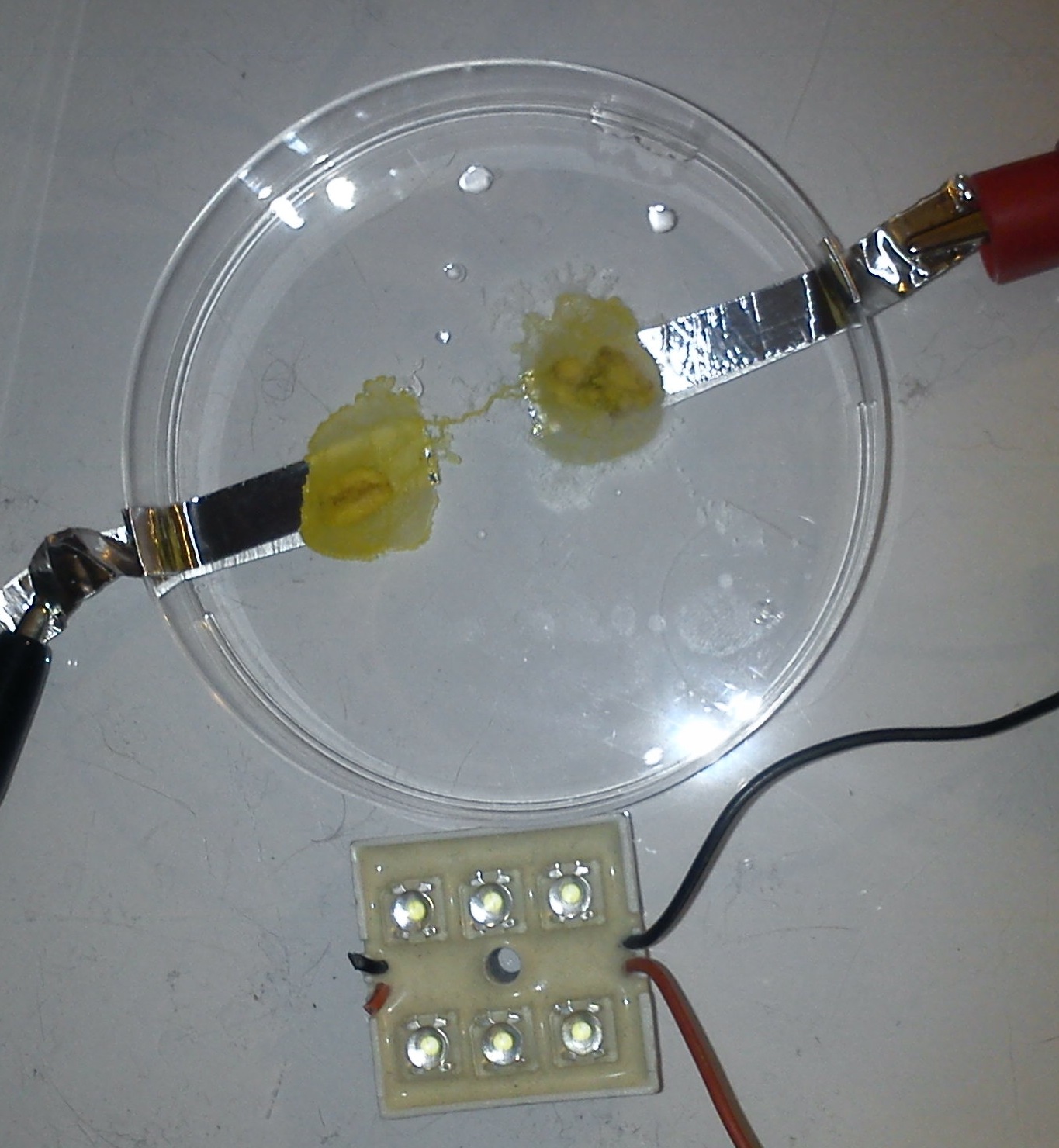}}
\subfigure[]{\includegraphics[width=0.49\textwidth]{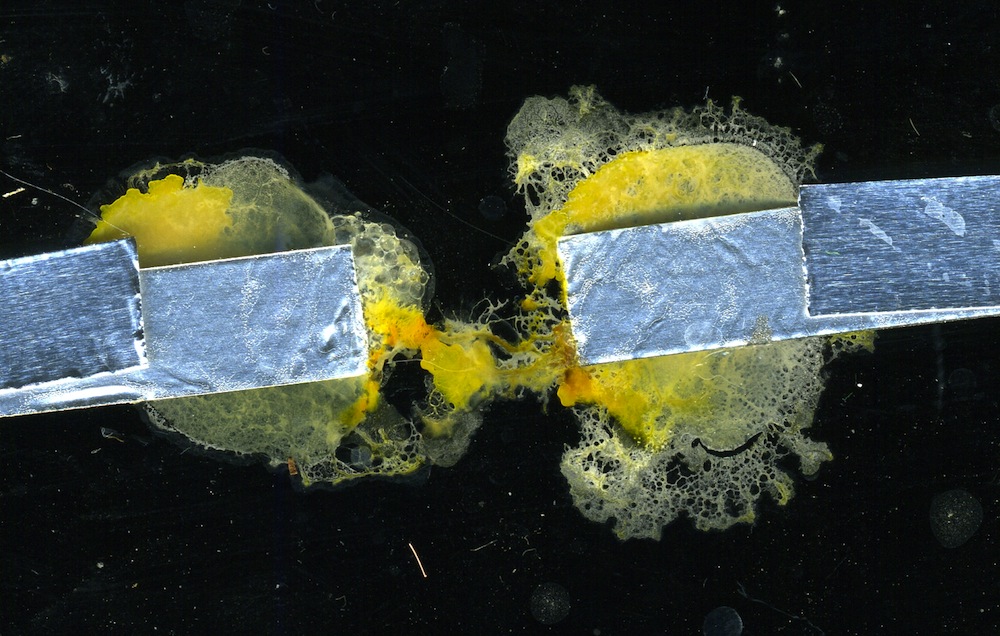}}
\caption{Physarum wire under load. (a)~Overall setup. 
(b)~Physarum wire 24~hr after functioning in the circuit with LED array.}
\label{LED}
\end{figure}

An undisturbed Physarum exhibits more or less regular patterns of oscillations of its surface electrical potential. 
The electrical potential oscillations are more likely controlling a peristaltic activity of protoplasmic tubes, necessary for distribution of nutrients in the spatially extended body of Physarum~\cite{seifriz_1937,heilbrunn_1939}. A calcium ion flux through membrane triggers oscillators 
responsible for dynamic of contractile activity~\cite{meyer_1979,fingerle_1982}.  It is commonly acceptable  that the potential oscillates with amplitude of 1 to 10~mV and period 50-200~sec, associated with shuttle streaming of cytoplasm~\cite{iwamura_1949, kamiya_1950, kashimoto_1958, meyer_1979}.  In our experiments we observed sometimes lower amplitudes because there are agar blobs between Physarum and electrodes and, also, electrodes were connected with a protoplasmic tube only. Exact characteristics of electric potential oscillations vary depending on 
state of Physarum culture and experimental setups~\cite{achenbach_1980}.  

In addition, low amplitude oscillations, in our experimental setup (two electrodes with agar blobs and Physarum on top) a potential difference between electrodes observed was usually 20-25~mV, and never exceeding 200~mV.  The Physarum potential even in extreme situations, is low. Thus by applying over 1~V to a protoplasmic tube we cause over-voltage. How does Physarum reacts to over-voltage, will its protoplasmic tube disintegrate? 
We have conducted 12 experiments on incorporating Physarum living wire into a circuit which includes an array of 6 LEDs (15~V white 10,000~Mcd).
In each experiment we applied c. 19~V to the circuit till the LED array was lighting bright. Potential on LED registered was 4.4-4.8~V and current 
 11-13~$\mu$A.  In all experiments Physarum wire was functioning for 24~h without loss of integrity (Fig.~\ref{LED}a). 

Input voltage stayed unchanged during 24~h yet after 24~h functioning as wires under load protoplasmic tubes decreased their resistivity: voltage registered on the LED array was 10.2~V and current running 3.9~$\mu$A. The decrease of resistivity is possibly due to increase of the overall mass of the Physarum wires, and, in some cases, growth of additional branches of tubes between the agar blobs. Typically,  
after one day of functioning under electrical load agar blobs start to dry out and Physarum gradually goes in the stage of 
sclerotisation (Fig.~\ref{LED}a).

\begin{figure}[!tbp]
\centering
\subfigure[]{\includegraphics[width=0.15\textwidth]{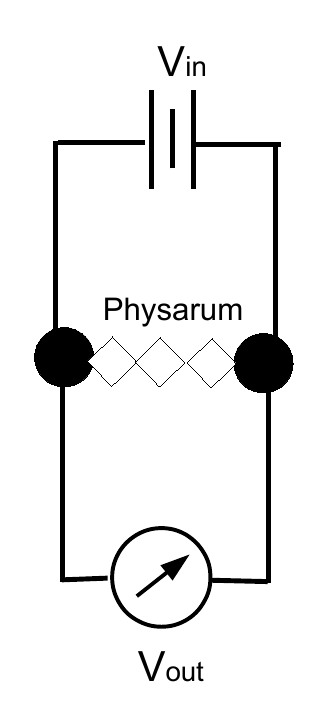}}
\subfigure[]{\includegraphics[width=0.7\textwidth]{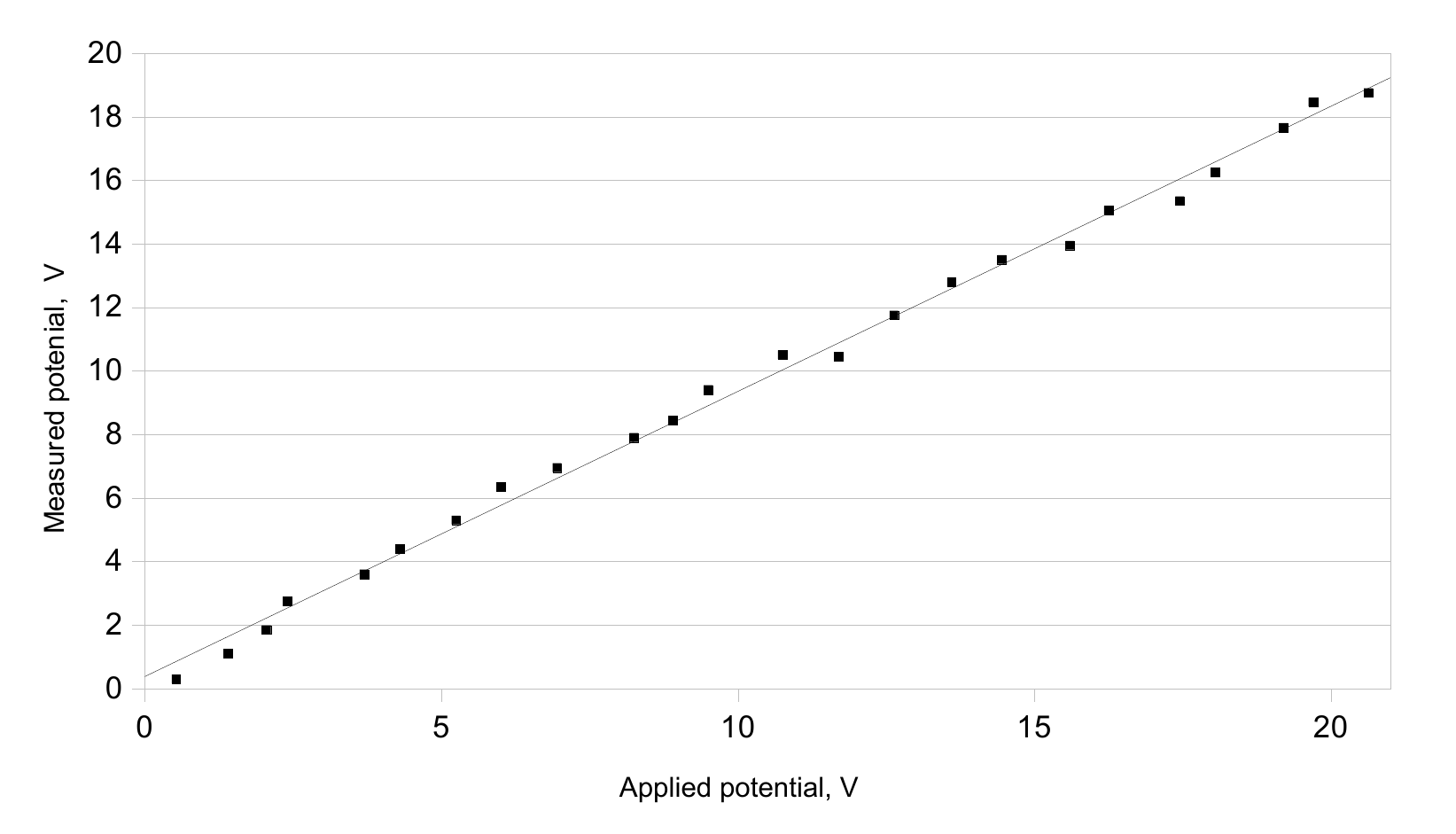}}
\subfigure[]{\includegraphics[width=0.7\textwidth]{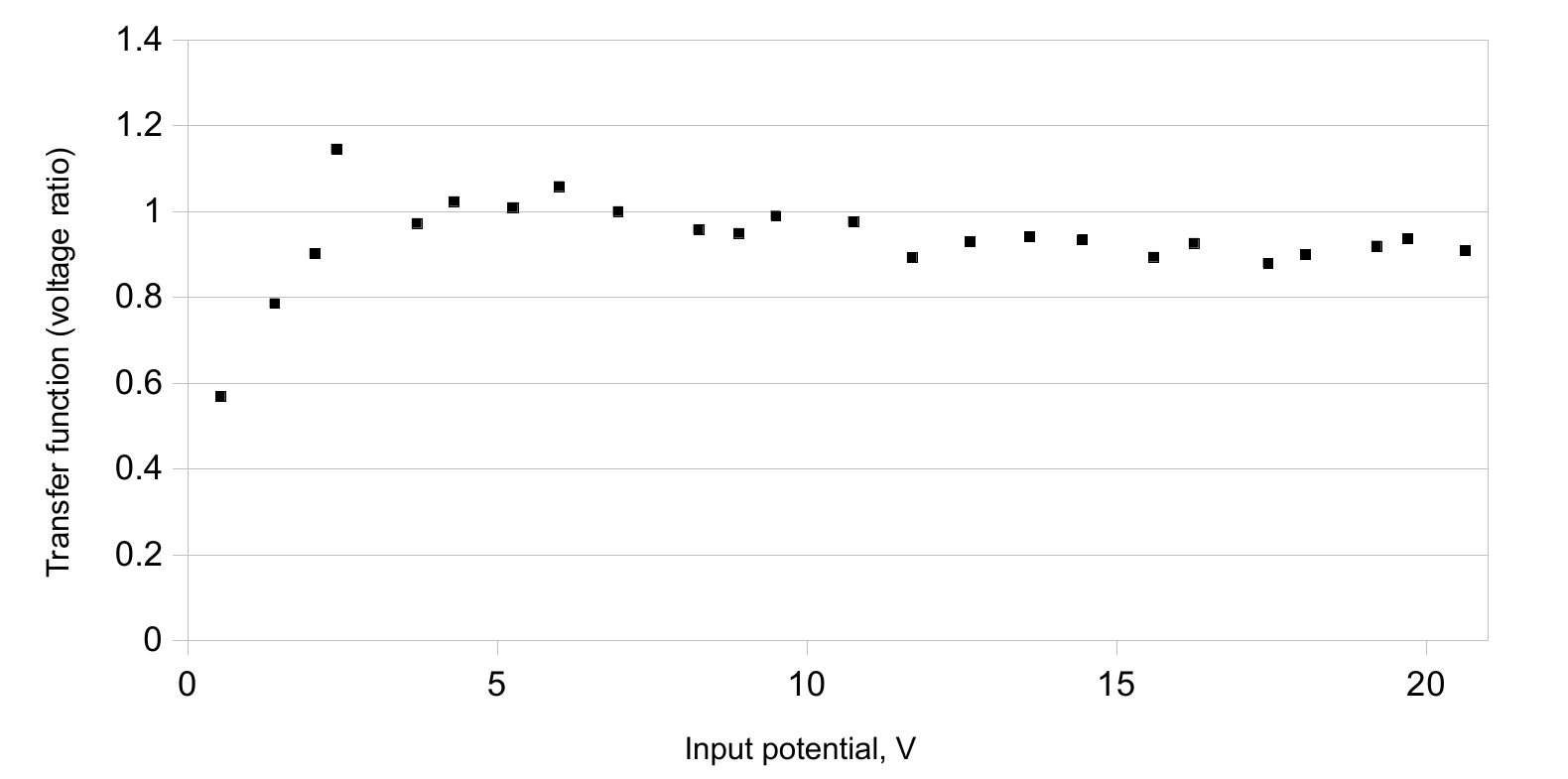}}
\caption{Potential dividing by Physarum impedances. 
(a)~Scheme of the experimental measurement. 
(b)~Input potential vs. output potential. 
(c)~Experimental plot of  a transfer function. 
}
\label{Vin2Vout}
\end{figure}

Potential $V_{out}$ recorded on a Physarum wire is a fraction of a potential $V_{in}$ applied to the 
Physarum wire. That is a Physarum acts as a potential divider:  see scheme of the circuit in Fig.~\ref{Vin2Vout}a.
The transfer function is linear (Fig.~\ref{Vin2Vout}b), subject to usual fluctuations of Physarum impedances in laboratory experiments
(Fig.~\ref{Vin2Vout}c).

\section{Routing Physarum wires}
\label{routing}

Growing Physarum circuits can be controlled by white~\cite{hader_1984} and blue~\cite{adamatzky_physarummachines} 
light, chemical gradients~\cite{knowles_1978, adamatzky_physarummachines}, temperature gradients~\cite{wolf_1997} 
and electrical fields~\cite{tsuda_2011}. In present section we evaluate routing in conditions far from an `idea experiment on 
slime mould taxis'. We grow and control Physarum grows on an almost bare breadboards.

\subsection{Routing with chemical fields} 

\begin{figure}[!tbp]
\centering
\subfigure[]{\includegraphics[width=0.505\textwidth]{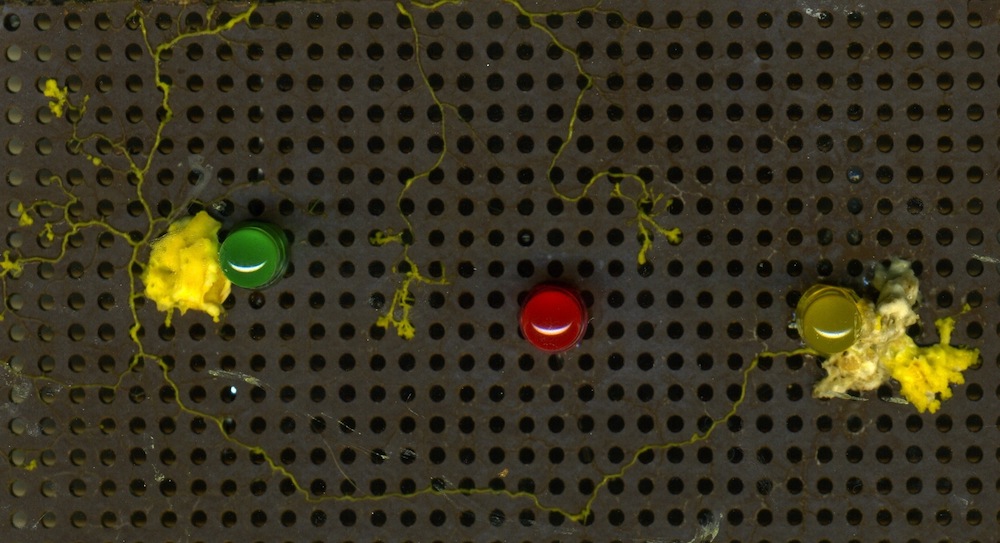}}
\subfigure[]{\includegraphics[width=0.485\textwidth]{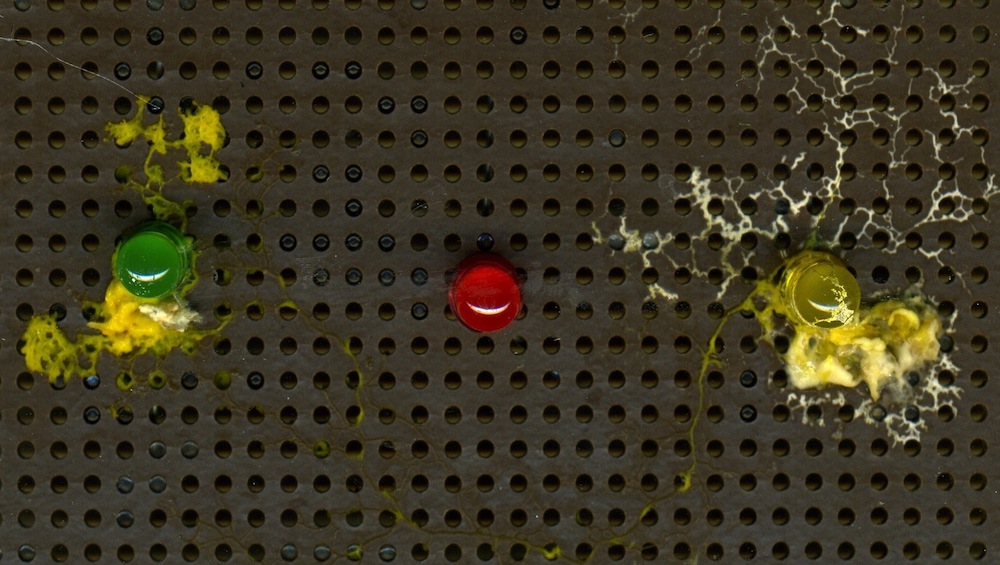}}
\subfigure[]{\includegraphics[width=0.6\textwidth]{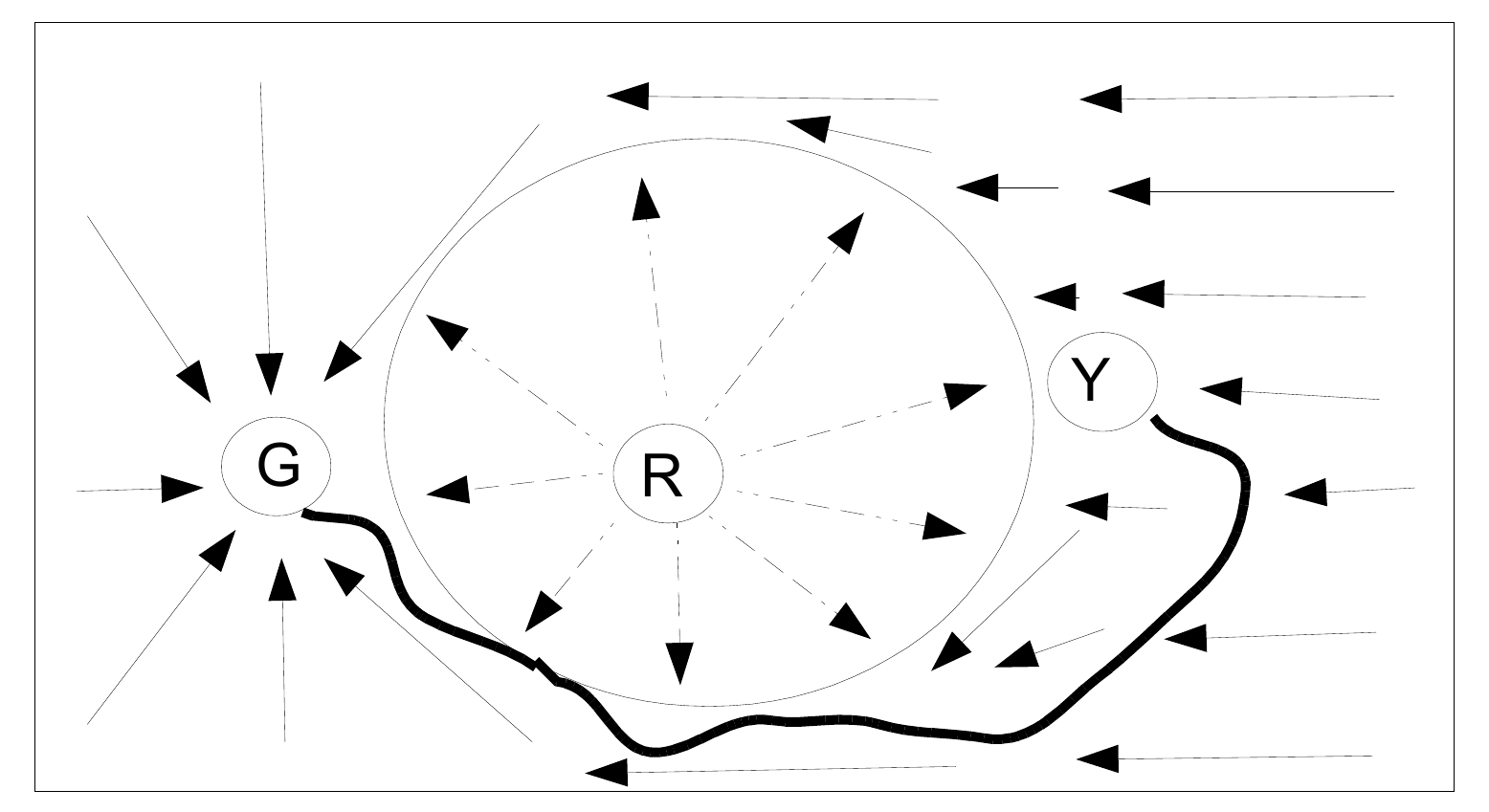}}
\caption{Snapshots of two experiments~(ab) on routing a Physarum wire  from yellow LED $Y$
to green LED $G$ with a constraint that a red LED $R$ must be avoided. Snapshots are made 36~h after
inoculation of Physarum. (c)~A scheme of attracting and repelling fields.}
\label{routingwithsalt}
\end{figure}

Results of two experiments are presented in Fig.~\ref{routingwithsalt}.  A task in both experiments 
was to route a Physarum wire from a position of yellow LED $Y$ to a position of green LED $G$ but avoiding 
a position of red LED $R$.   In both experiments Physarum was inoculated nearby $Y$. An oat flake 
was placed east of $G$. Chemoattractants released either by the oat flake or by bacteria colonising 
it diffused in the air and attracted Physarum. To prevent the slime mould going nearby $R$ we placed 
a grain of salt near $R$. The salt absorbed water from humid environment of the experimental setup
and diffused in the agar layer underlying the breadboard.  Physarum is repelled by high concentration of 
salt in its substrate. Thus the Physarum moved towards attracting $G$ and, at the same time, 
avoided repelling $R$ (Fig.~\ref{routingwithsalt}c). 

After the Physarum wire was established and got into direct contact with pins of $Y$ and $G$ we measured an 
electrical potential between the pins. Two experiments are illustrated in Fig.~\ref{routingwithsalt}.  In experiment 
shown in  Fig.~\ref{routingwithsalt}a) a protoplasmic tube connecting the $Y$ and $G$ exhibited potential 37-40~mV, 
and in experiment shown in Fig.~\ref{routingwithsalt}b the tube's potential was 40-45~mV. These conform well with 
natural variance of potential in protoplasmic tubes of \emph{P. polycephalum}.  Resistivity measured between the pins 
was 1300-1500~$\Omega \cdot$cm,  in a range of electrical resistivity of biological tissue~\cite{geddes_baker_1967}.

\begin{figure}[!tbp]
\centering
\includegraphics[width=0.6\textwidth]{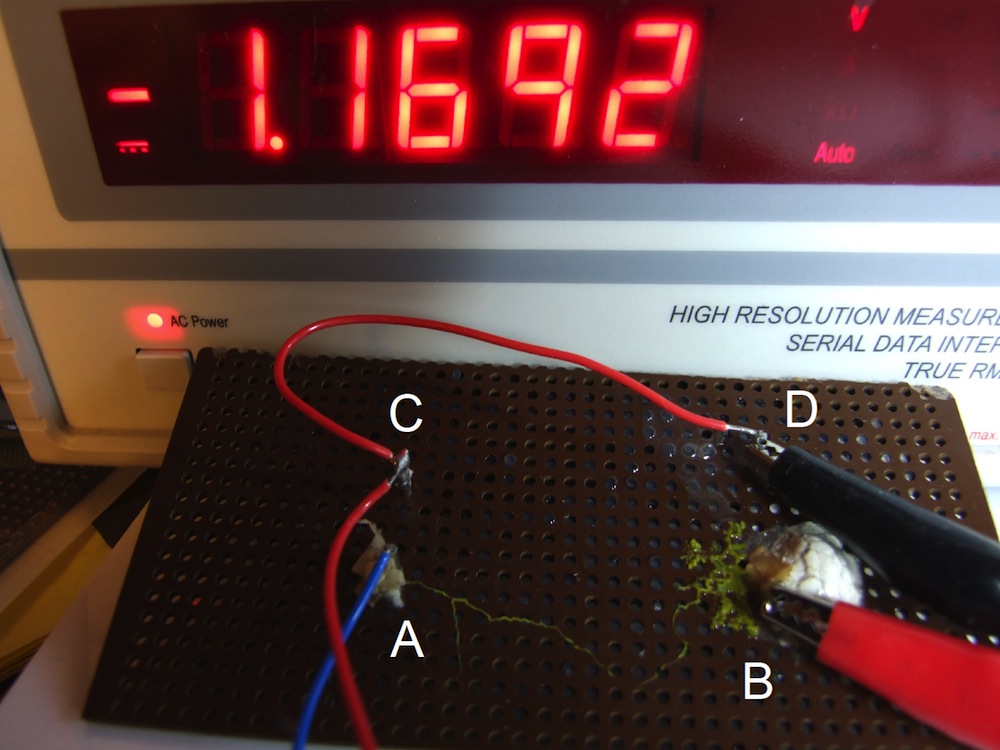}
\caption{Routing Physarum wires with somniferous pills.}
\label{board1}
\end{figure}

An example of routing with non-food chemo-attractants is shown in Fig.~\ref{board1}.  To complete a circuit we needed to grow
 a Physarum wire between points 'A' and 'B'. We inoculated Physarum so it was in a direct contact with pin 'A'  and placed half a pill 
 of a valerian-containing herbal remedy Kalms, see details in~\cite{adamatzky_tablets}, nearby pin 'B'.  Being attracted to valerian 
 Physarum propagated from pin 'A' to pin 'B" and forms a conductive pathways between the pins. A difference of a membrane potential of the 
 protoplasmic tube between 'A' and 'B' sites was in a range of 50~mV. When 8.5~V DC applied to inputs 'A' an 'C' of the hybrid circuit (Fig.~\ref{board1}) a potential slightly below 1.2~V was recorded between pins 'D' and 'B'.
 
 \begin{figure}[!tbp]
\centering
\subfigure[]{\includegraphics[width=0.49\textwidth]{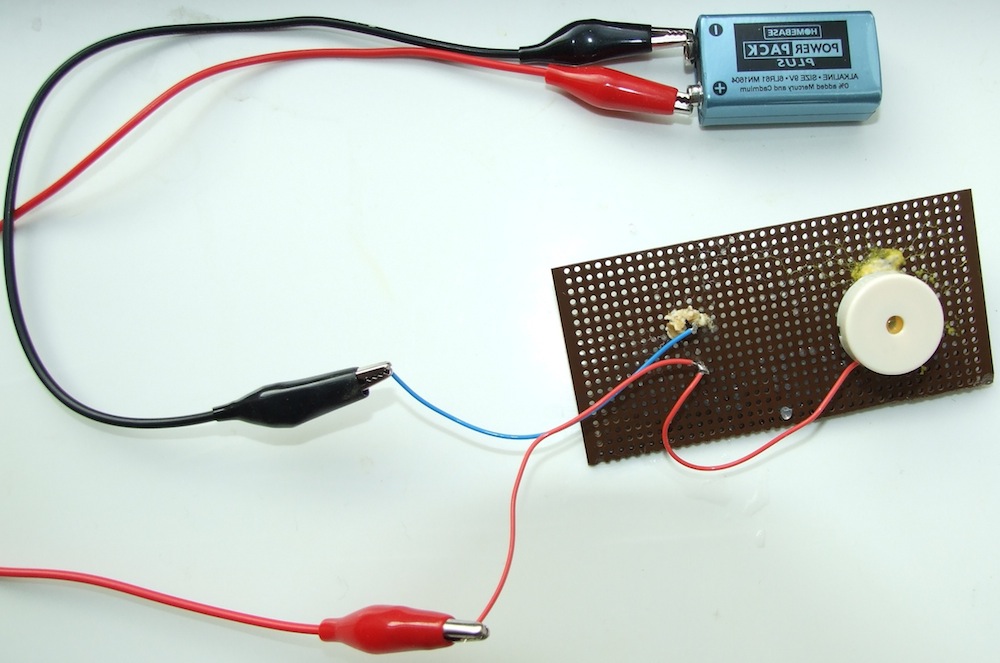}}
\subfigure[]{\includegraphics[width=0.49\textwidth]{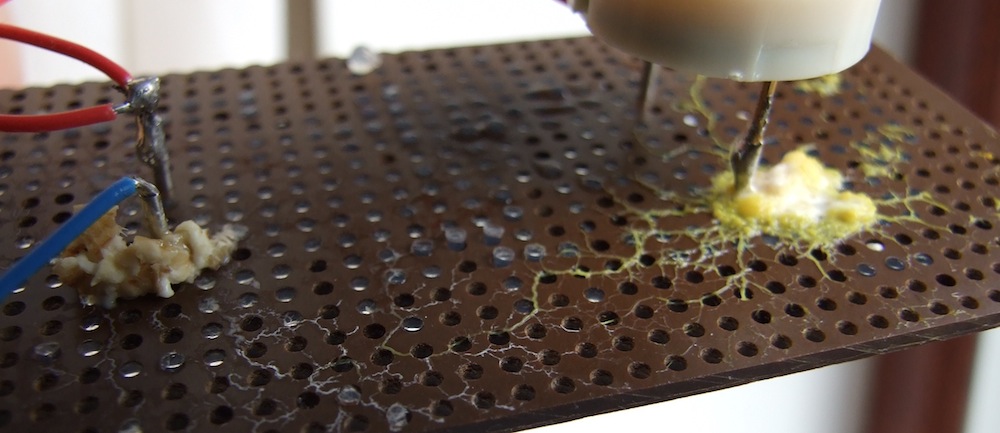}}
\caption{Functionality of Physarum wires routed with a valerian-containing pill.
(a)~Overview of the circuit. 
(b)~A Physarum wire connects pin battery's anode to a pin of a piezo transducer.}
\label{boardbuzzer}
\end{figure}

In an experiment illustrated in Fig.~\ref{boardbuzzer} Physarum developed a  conductive pathway in a circuit including a DC voltage supply and 
a piezo audio transducer (Kingstate KPEG158-P5, operational for 1-20~V, current consumption max 7~A). 
On applying 8~V DC to inputs of the circuit we registered 2.2~V potential on the piezo transducer's pins.  The transducer 
produced sound near 30~dB for a minimum duration of 10~min. Physarum wire remained alive and did not change its morphology during 
the circuit's operational mode.

\subsection{Routing with electric fields}

\begin{figure}[!tbp]
\centering
\subfigure[]{\includegraphics[width=0.49\textwidth]{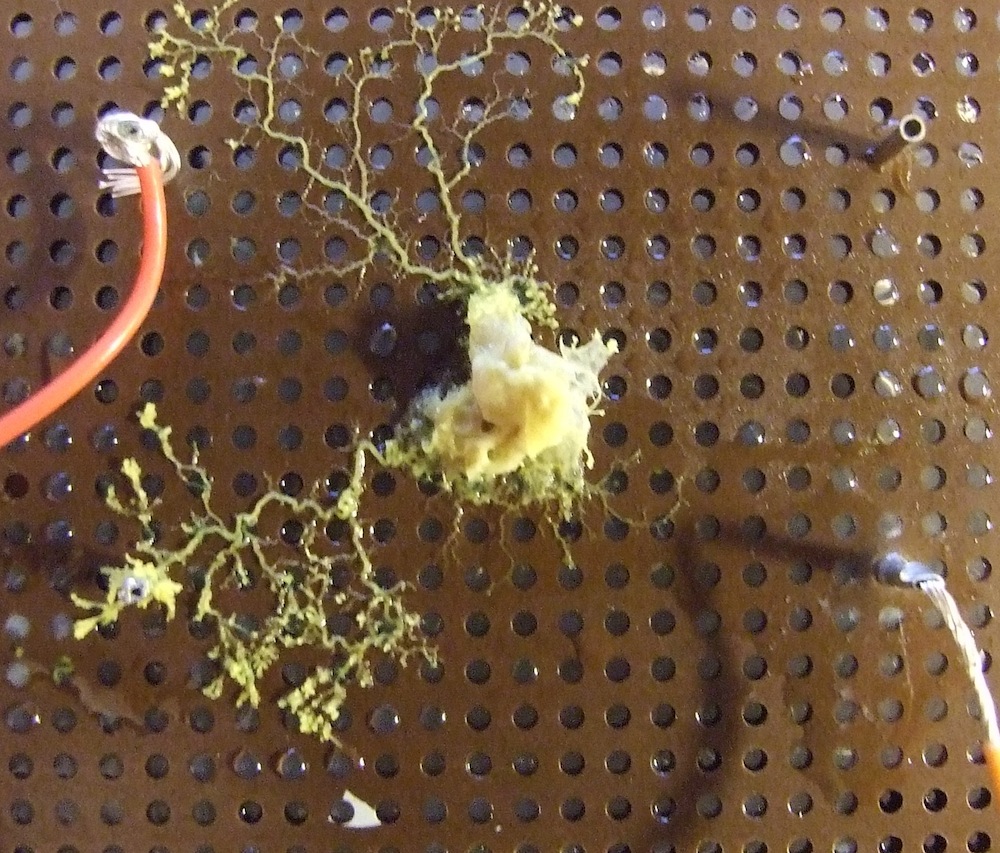}}
\subfigure[]{\includegraphics[width=0.49\textwidth]{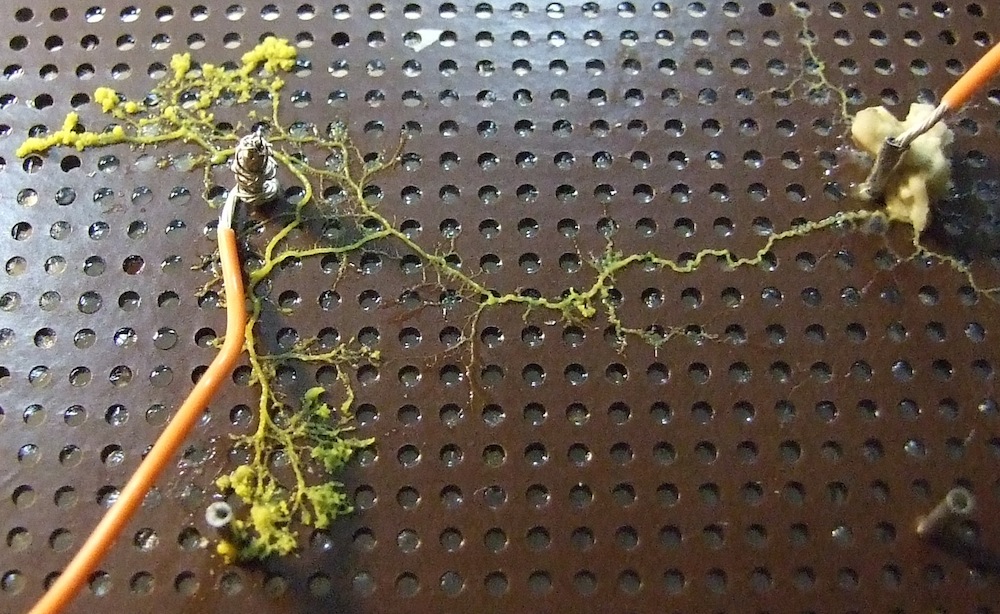}}
\subfigure[]{\includegraphics[width=0.7\textwidth]{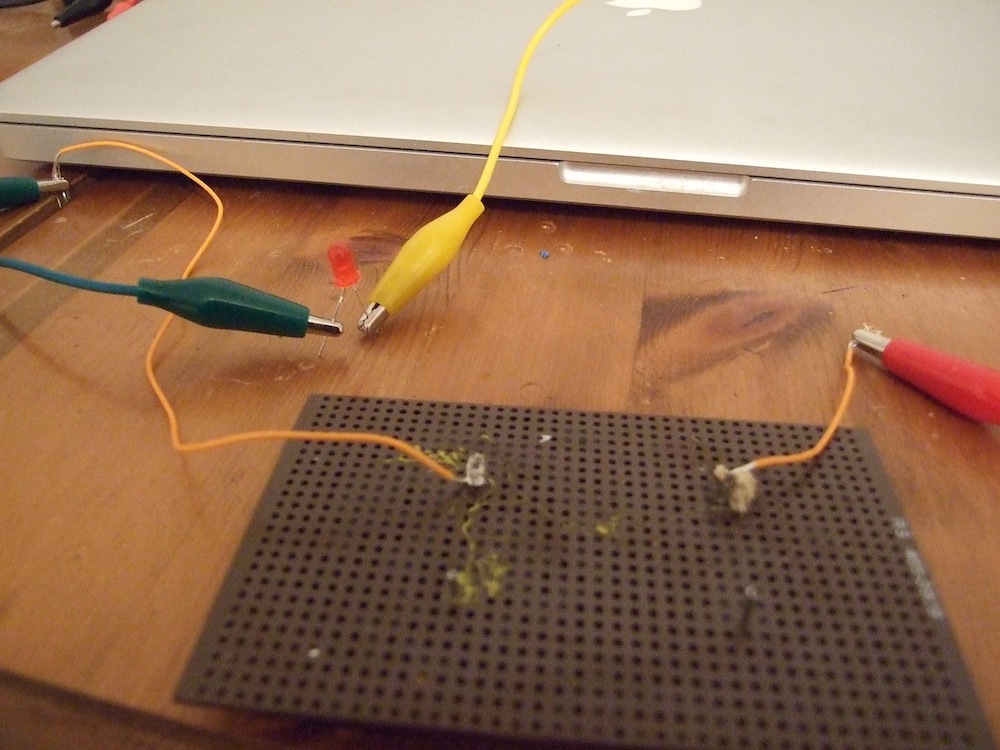}}
\caption{Routing Physarum wires with electro-magnetic fields. 
(ab)~Snapshots of two experiments on routing Physarum wire with EMF.  An 1.6V battery is connected to two pins on the board. 
Cathode is on the left. Physarum propagates towards cathode. (c)~A hybrid circuit including Physarum wire is used to light up a LED.}
\label{battery}
\end{figure}

Controlling growth of Physarum with chemo-attractants and repellents is proved to be a reliable method of routing living wires. 
However, direct contact of attractants/repellents with growth and Physarum itself is undesirable: the substrate might become
contaminated with the chemicals preventing further re-routing of the wires. Also, chemicals diffusing in a substrate might affect 
functioning of existing Physarum wires. Non-invasive control techniques, as e.g. with light~\cite{adamatzky_2009} are appropriate
yet unreliable and do not always give a predictable control of the growing slime mould. The Physarum plasmodium is known to shown negative galvanotaxis: the slime mould migrates towards cathode under an electric field~\cite{anderson_1951, anderson_1962}.  A fine pattern 
of Physarum growth can be achieved by  culturing the slime mould on an array of probes, where direction, migration and obstacle avoidance
behaviour of Physarum could be guided by the spatial distribution of electric currents formed by the probes~\cite{tsuda_2011}.

To evaluate galvanotaxis based routing of Physarum wires in a conditions more close to realities of circuit design we undertook
15 experiments with Physarum wires growing on a breadboard with a limited agar substrate: agar plates were fixed underneath the 
breadboards, thus Physarum was only able to contact agar gel through the holes in the breadboards. In each experiment we 
placed 2-3 oat flakes colonised by Physarum onto a breadboard with several pins. Two of the pins were connected
to a 1.6~V battery. 

We found that in 10 of 15 experiments the slime mould propagated to cathode. Two experiments are illustrated in Fig.~\ref{battery}.  
In experiment Fig.~\ref{battery}a Physarum was inoculated between the pins. In experiment Fig.~\ref{battery}b the slime mould was placed nearby anode. In both experiments controllable propagation to cathode is recorded.  Conductivity of Physarum wire was sufficient to light up a LED 
by applying 9~V DC to the circuit (Fig.~\ref{battery}c).

\section{Self-repair of Physarum wires}
\label{self_repair}

\begin{figure}[!tbp]
\centering
\subfigure[]{\includegraphics[width=0.49\textwidth]{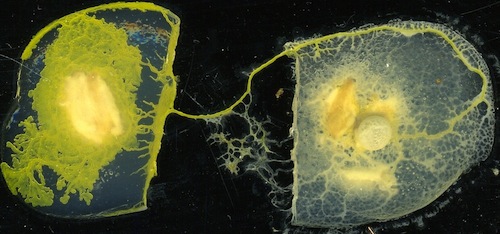}}
\subfigure[]{\includegraphics[width=0.49\textwidth]{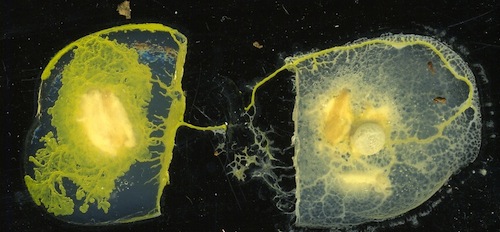}}
\subfigure[]{\includegraphics[width=0.49\textwidth]{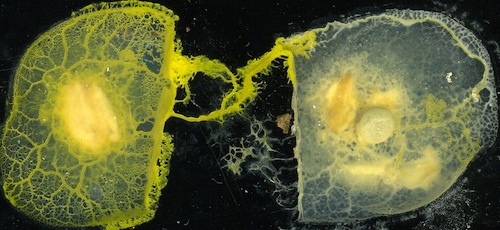}}
\subfigure[]{\includegraphics[width=0.9\textwidth]{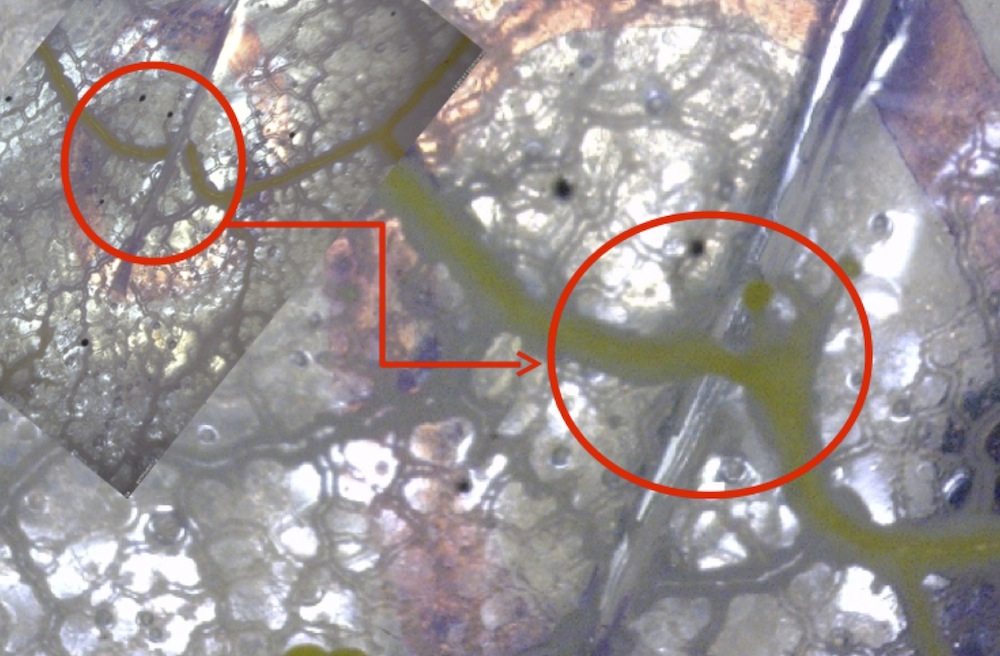}}
\caption{Response of a Physarum wire to cutting: (abc)~protoplasmic tube self-restores its integrity while on a bare plastic substrate,
and (d)~ends of cut tube almost exactly merge while tube is on an agar gel.}
\label{selfrepairscans}
\end{figure}

Physarum wires can self-repair after a substantial damage.  In 12 experiments we found that after part of a protoplasmic tube 
is removed (1-2~mm segment) the tube restores its integrity in 6-9~hr.  Typically, a cytoplasm from cut-open ends spills out on a substrate. 
Each spilling of cytoplasm becomes covered by a cell wall and starts growing.  In few hours growing parts of the tube meet with each other 
and merge. 

An example of tube's self-repair is shown in Fig.~\ref{selfrepairscans}a--c.  In Fig.~\ref{selfrepairscans}a we see an undamaged tube 
connecting two agar blobs. The tube propagates on a bare plastic substrate. A segment of the tube is removed, see 
Fig.~\ref{selfrepairscans}b. In approximately 8~hr after the damage the tube restores its integrity, see Fig.~\ref{selfrepairscans}c. 

Growing ends of a damaged tube not necessarily meet up each other exactly when developing on a bare plastic substrate. However, exact
merging of the ends is almost always the case when tube is resting on an agar substrate. See example in Fig.~\ref{selfrepairscans}d.
In 2-3~hr after a tube was cut (top left insert in Fig.~\ref{selfrepairscans}d) active growth zones are formed at the ends of the tube. 
They propagate towards each other and merge (main photo in Fig.~\ref{selfrepairscans}d). 

\begin{figure}[!tbp]
\centering
\includegraphics[width=0.95\textwidth]{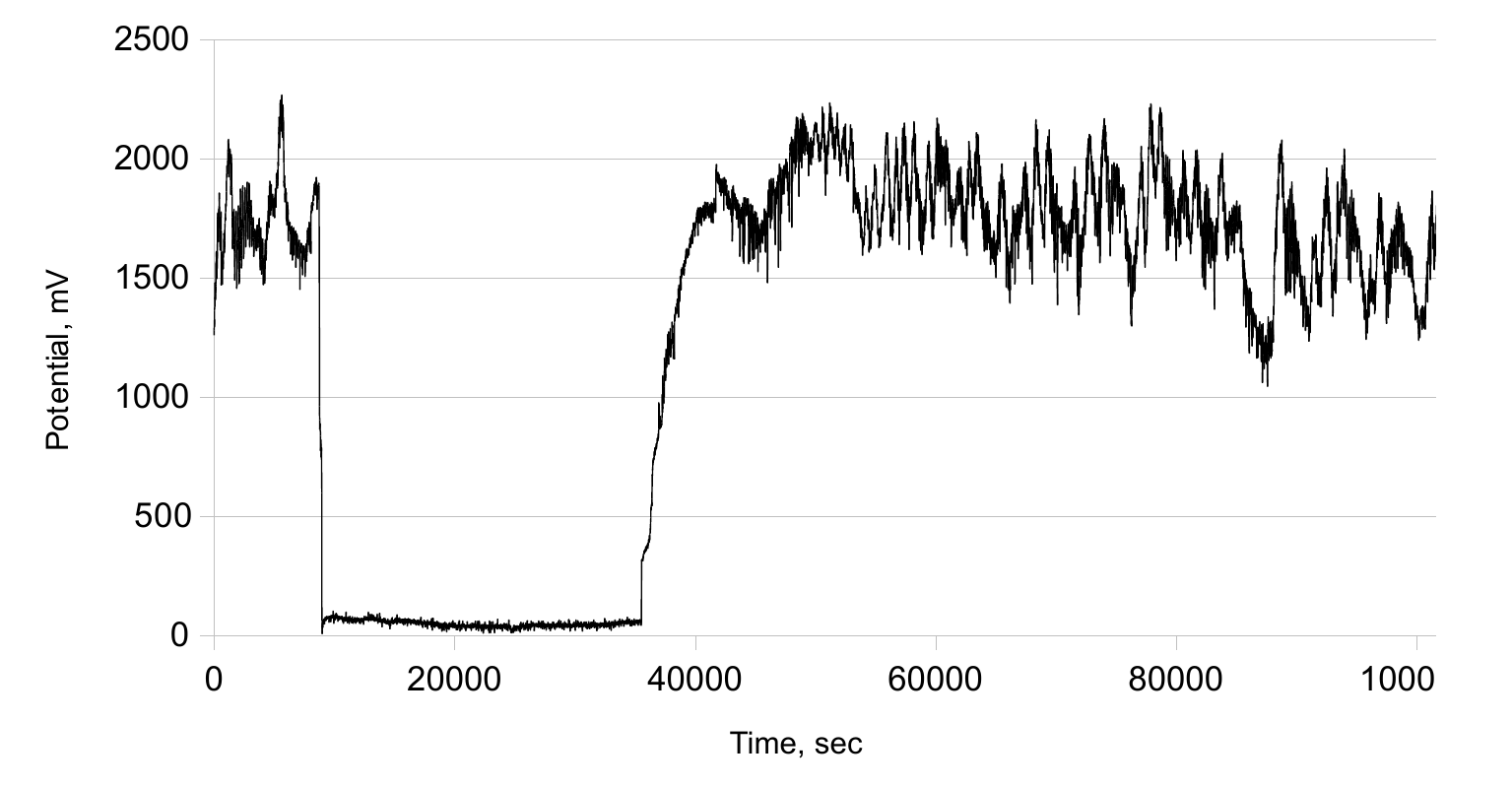}
\caption{Response of a Physarum wire to cutting. Horizontal axis is a time in sec, vertical axis is voltage in mV. When Physarum wire is damaged 
voltage drops to zero, when the wire self-heals the voltage restores to its original value.}
\label{selfrepairpotential}
\end{figure}

Restoration of tubes conductivity as a result self-repair was confirmed by electrical measurements, see 
Fig.~\ref{selfrepairpotential}.  A potential 4~V DC was applied to a Physarum wire (10~mm length protoplasmic tube as in all 
previous setups). Under the applied voltage the wire showed average potential 1.9~V, moving between 1.5~V 
to 2.3~V.  150~min after start of recording we destroyed parts of the wire: 1.8~mm of protoplasmic tube was removed from each wire. 
The wire ceased to be conductive. This was reflected in a sharp voltage drop (Fig.~\ref{selfrepairpotential}).

Several hours after being cut the protoplasmic tube self-healed and formed a fully functioning tube again.  The Physarum wire started to return to its conductive state 440~min after being damaged and returned to its fully conductive state in next 230~min: it shows 1.8~V, oscillating between 1.6~V abd 2.1~V when 4~V DC is applied  (Fig.~\ref{selfrepairpotential}).

\section{Propagating on electronic boards}
\label{electronic_boards}

\begin{figure}[!tbp]
\centering
\subfigure[]{\includegraphics[width=0.45\textwidth]{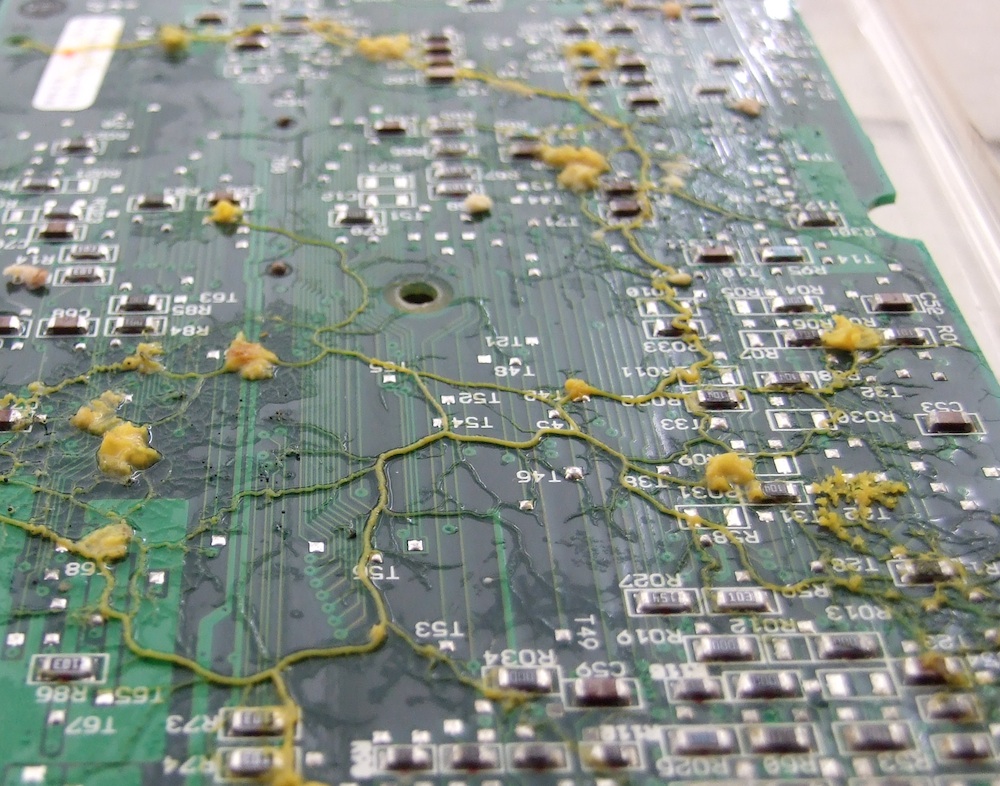}}
\subfigure[]{\includegraphics[width=0.45\textwidth]{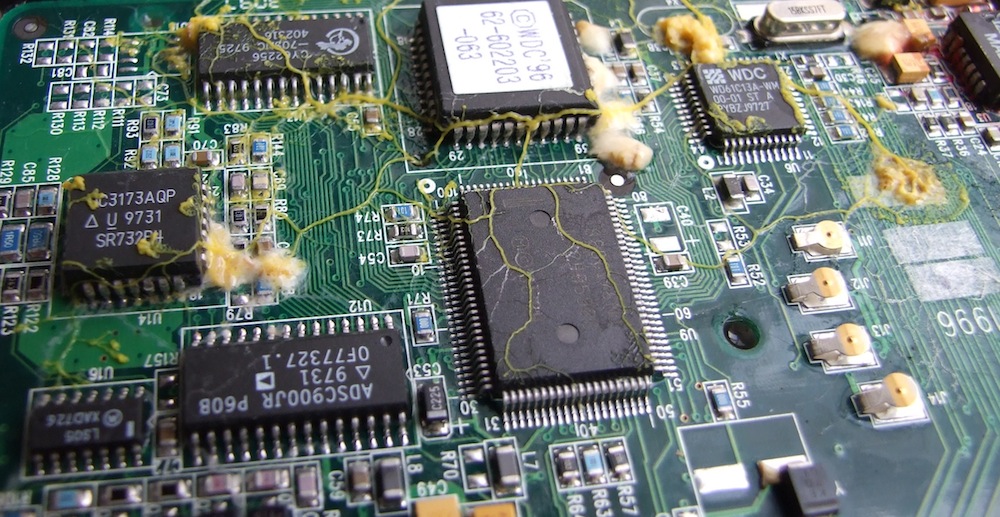}}
\subfigure[]{\includegraphics[width=0.49\textwidth]{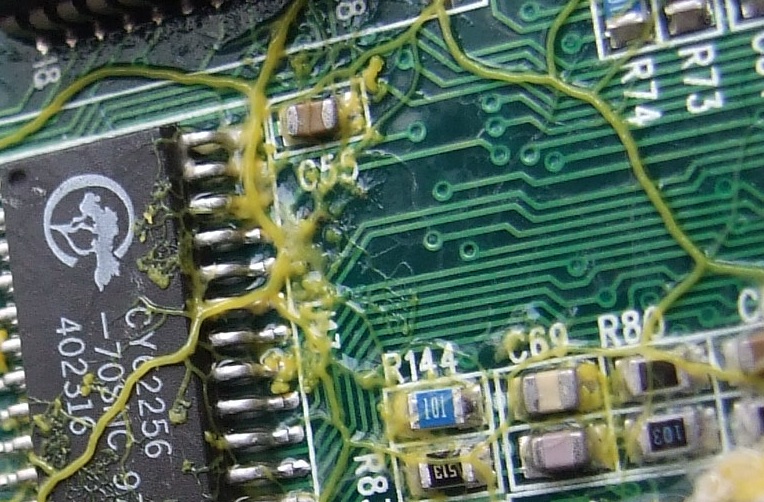}}
\caption{Photographs of hard disc interface cards with slime mould growing on top.
(a)~Overview of the board with Physarum.
(b)~Slime mould propagates on the board's elements.
(c)~Enlarged view of a board, colonised by Physarum, to demonstrate 
that width of a protoplasmic is compatible with conductive pathways on the board.
}
\label{boards}
\end{figure}

To evaluate how well Physarum propagates on a bare surface of electronic components and assemblies we
conducted experiments with electronic boards (Fig.~\ref{boards}). Typically, a slime mould was inoculated at one edge of the board, 
e.g. further edge in Fig.~\ref{boards}a, and oat flakes were scarcely distributed on the board to attract Physarum to certain domains of 
the boards. The oat flakes generated chemo-attractive fields to guide Physarum wires towards imaginary pins. The boards were kept in a container, with a shallow water on the bottom to keep humidity very high. The boards with Physarum were not in direct contact with the water.

The slime mould propagated on the boards with a speed of between 1~mm to 5~mm per hour. Physarum propagated satisfactory on 
both sides of the boards, and usually spanned a planar set of oat flakes with networks of protoplasmic 
tubes ranging from spanning trees to their closures into  $\beta$-skeletons (Fig.~\ref{boards}ab).  As shown in Fig.~\ref{boards}c
a width of protoplasmic wires grown by Physarum is comparable with a width of conductive pathways on the computer boards.

\section{Insulating Physarum wires}
\label{insulating}

\begin{figure}[!tbp]
\centering
\subfigure[]{\includegraphics[width=0.49\textwidth]{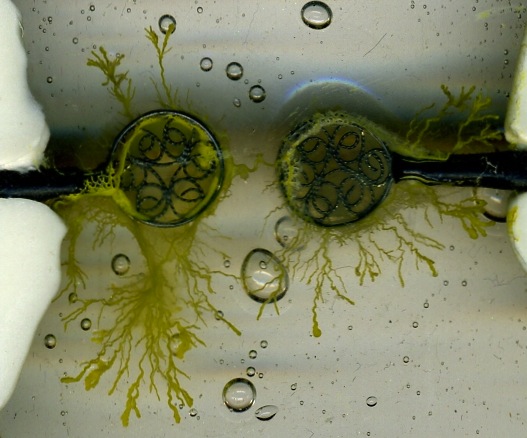}}
\subfigure[]{\includegraphics[width=0.49\textwidth]{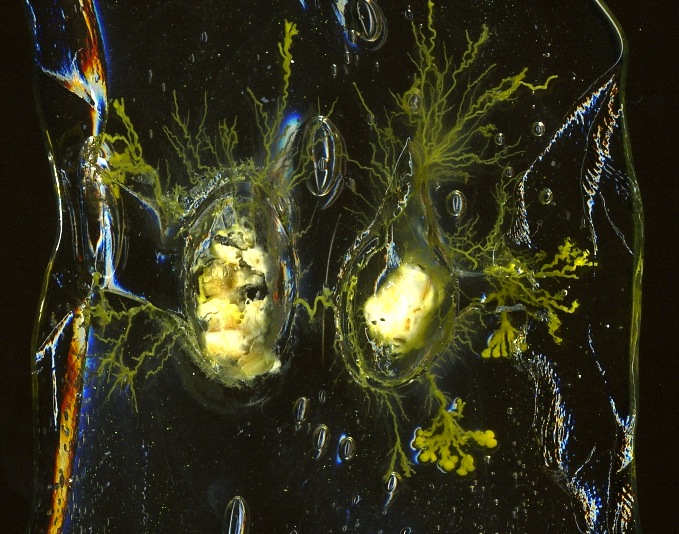}}
\subfigure[]{\includegraphics[width=0.7\textwidth]{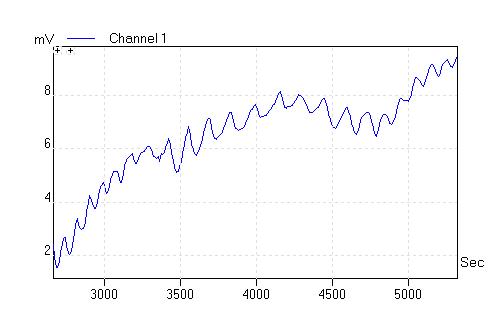}}
\subfigure[]{\includegraphics[width=0.5\textwidth]{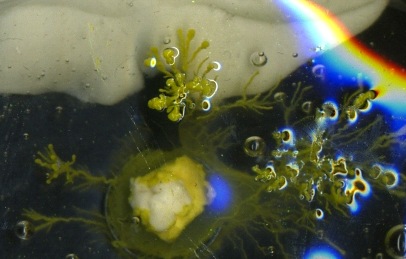}}
\caption{Insulating Physarum wires. 
(a)~Physarum wire, connecting two probes is covered by a layer of D4; view from the bottom of the Petri dish. 
(c)~Physarum wire in a layer of D4 continues functioning when removed from the Petri dish. 
(c)~Recording of electrical potential of Physarum between two probes, as in (a), during 83~min. 
(d)~'Breathing' of Physarum.}
\label{physarumgum}
\end{figure}

We successfully tested insulation of Physarum wires with octamethylcyclotetrasiloxane 
(Silastic 4-2735 Silicone Gum, Dow Corning S.A., B-7180 Seneffe, Belgium). We will use name D4 for brevity. 
The D4 is a liquid at normal temperature and pressure; its melting point is around 17-18~C$^o$~\cite{merk_1996}.
Its density is about 0.96 g/cm$^3$~\cite{merk_1996}. In experiments we gently poured D4 over Petri dish with probes 
and Physarum until a layer of D4 covered all objects on the bottom of the Petri dish (Fig.~\ref{physarumgum}a). 

Physarum becomes completely encapsulated in the silicon insulator. When a slab of D4 removed from the Petri dish Physarum 
remains inside  (Fig.~\ref{physarumgum}b). Physarum survived inside D4 for hours. Figure~\ref{physarumgum}c shows electrical activity of Physarum 
2 hours after it was immersed into D4: the slime mould exhibits 'textbook classical' oscillations of its electrical potential with 
amplitude around 1~mV and period 145~sec. Coincidently, the width of D4 cover is thinner on top of probes. There Physarum forms 
vertically ascending protrusions which reach surface of the D4 and thus Physarum is capable for intake oxygen and keeping the 
protoplasmic tube connecting probes alive (Fig.~\ref{physarumgum}d).

\section{Discussions}
\label{discussion}

\begin{figure}[!tbp]
\centering
\includegraphics[width=0.9\textwidth]{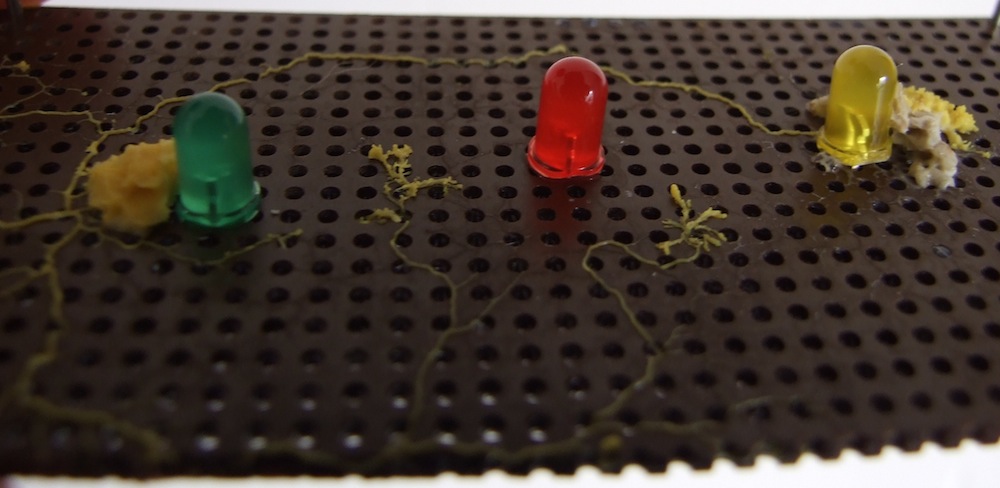}
\caption{ Example of undesired sprouting. Physarum is always in the flux, and additional wires can grow.}
\label{physarumflux}
\end{figure}

Growing wires from slime mould is a perspective direction of research in bio-inspired novel computing substrates. In experimental
laboratory studies we shown that 
\begin{itemize}
\item Physarum's protoplasmic tubes remains functioning and conductive under reasonable high load and capable to act as wires (Physarum wires)
in electrical circuits including lightning and actuating devices. 
\item Physarum wires can be routed on various types of substrates using chemo-attractants, chemo-repellents and 
electromagnetic fields. 
\item Physarum wires can be insulated with silicon. The wires remain alive and functioning while covered in insulator for days.
\item Physarum can self-repair. When a Physarum wire is cut the damaged ends grow towards each other and merge, 
thus restoring integrity of the original wire.
\end{itemize}

Using living protoplasmic tubes as wires suffer from few disadvantages. Physarum is always in motion and newly developed protoplasmic tubes can 
interfere with existing tube-wires. For example, in experiment shown in Fig.~\ref{physarumflux} Physarum grown a protoplasmic
wire between two LEDs: the wire is visible near the farthest edge of the board. Few hours after the wire formed, the 
slime mould continued its foraging behaviour and sprouted two undesired protoplasmic tubes. 
These tubes grown near the closest edge of the board as seen in Fig..~\ref{physarumflux}. We should find reliable ways of 
inhibiting sprouting after all required wires are formed.

\begin{figure}[!tbp]
\centering
\subfigure[]{\includegraphics[width=0.6\textwidth]{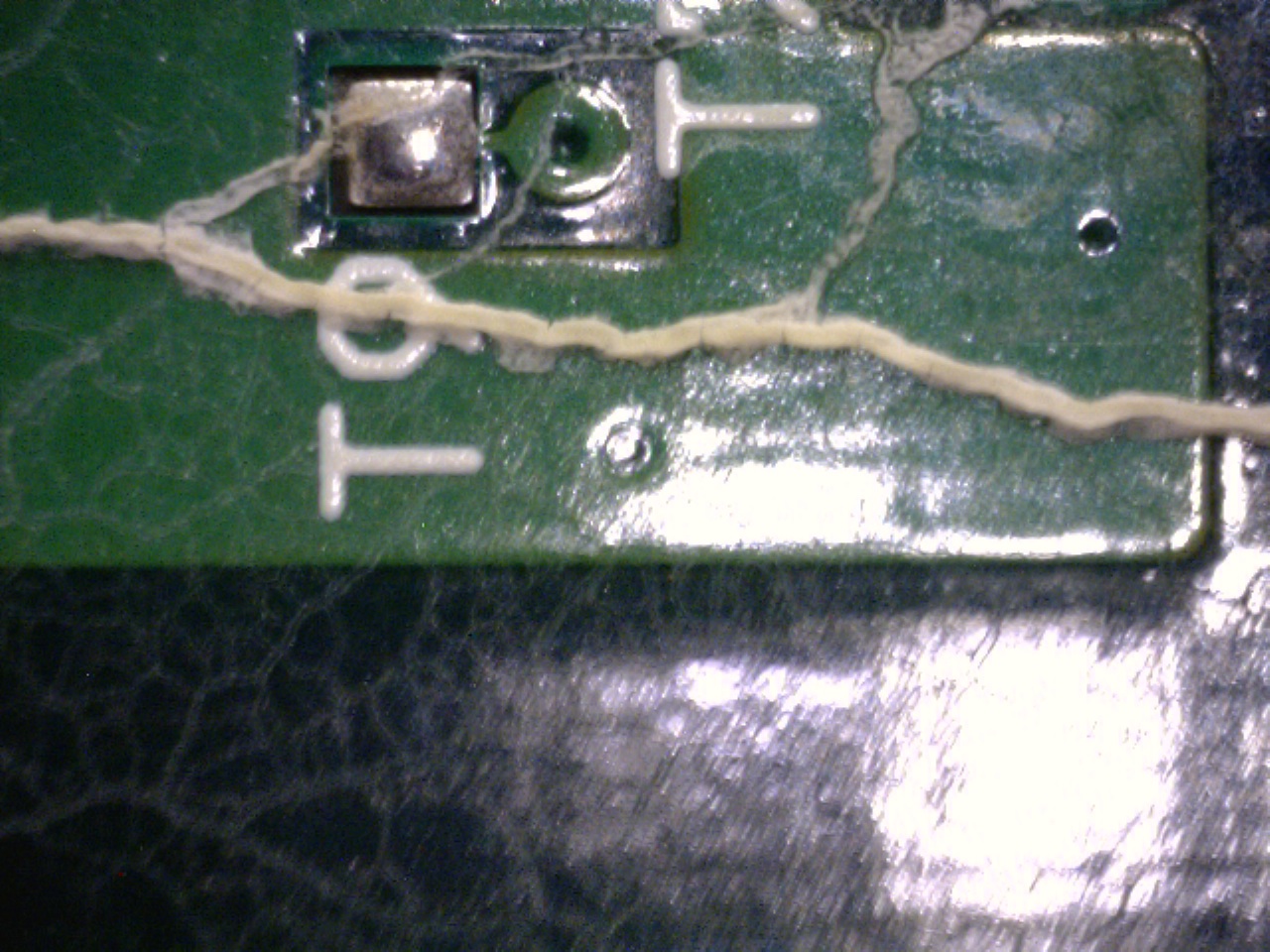}}
\caption{Abandoned protoplasmic tube still keeps its shape preserved. 
If loaded with conductive beads it would play a role of permanent wire. }
\label{abandoned}
\end{figure}

In present paper we considered living Physarum wires. Their life time is limited. Typically in 5-7 days, may be longer, depending on humidity, a 
protoplasmic tube becomes abandoned   (Fig.~\ref{abandoned}).  After becoming abandoned the tube keeps, up to some degree,  
its water contents, and thus conductivity, for few more days and then dries up and its walls collapse. 

Physarum wires are volatile.  Circuits with the protoplasmic wires can function for days but after a week at most Physarum might migrate away, 
or get into a sclerotium phase (if low humidity) or fructify (if exposed to light) or just get colonised by some other moulds and vanish. How to make Physarum wires long-lasting?  When the slime mould develops a network of protoplasmic tubes spanning sources of nutrients, the cell maintains its integrity by pumping nutrients and metabolites between remote parts of its body via cytoplasmic 
streaming~\cite{Stewart_Stewart_1959, Allen_1963,  Newton_1977, Hulsmann_Wohlfarth-Bottermann_1978}. The cytoplasmic streaming could be employed for the transportation of bio-compatible substances inside the protoplasmic network. In \cite{Adamatzky_2010a} we demonstrated that the plasmodium of \emph{P. polycephalum} consumes various coloured dyes and distributes them in its protoplasmic network. By specifically arranging a configuration of attractive (sources of nutrients) and repelling (sodium chloride crystals) fields we can program the plasmodium to implement the following operations: to take in specific coloured dyes from the closest coloured oat flake; to mix two different colours to produce a third colour; and, to transport colour to a specified locus of an experimental substrate. Transportation of colourings \emph{per se} is of little interest but shows the potential of \emph{P. polycephalum} as a programmable transport medium. To employ the slime mould's potential of internalisation and re-distribution of foreign particles for the development of a long-life wires we must consider suitable functional materials. In~\cite{mayne_2013}, 
inspired by our previous results \cite{Adamatzky_2010a} and studies on cellular endocytosis of magnetic nano-beads \cite{Li_2005} and fluorescent nano-beads \cite{Bandmann_2012},  and nanowire scaffolding for living tissue \cite{Tian_2012}, we demonstrated internalisation and transport of  two types of functional materials: magnetic nano-particles and glass spheres coated with silver. The particles could be redistributed inside the Physarum body and indeed 'spread' along protoplasmic tubes. Initial results, rather a proof of loading, are reported by us in~\cite{mayne_2013}. Final adjustments to  a distribution of magnetic particles can be controlled using magnetic twizzers~\cite{barbic_2002}. Mineralisation of grown Physarum wires to make them permanent wires will be a continued topic of further studies.

\end{document}